\documentclass[usenatbib]{mn2e}
\usepackage{graphicx,color,epsfig}

\newcommand{\apj}{ApJ}
\newcommand{\apjs}{ApJS}
\newcommand{\mnras}{MNRAS}
\newcommand{\icarus}{ICARUS}
\newcommand{\aap}{A\&A}
\newcommand{\araa}{ARA\&A}
\newcommand{\apjl}{ApJL}

\newcommand{\nat}{Nature}

\def\ltsima{$\; \buildrel < \over \sim \;$}
\def\simlt{\lower.5ex\hbox{\ltsima}}
\def\gtsima{$\; \buildrel > \over \sim \;$}
\def\simgt{\lower.5ex\hbox{\gtsima}}

\def\msun{{\,{\rm M}_\odot}}

\newcommand\mearth{{\,{\rm M}_{\oplus}}}

\newcommand\mj{{\,{\rm M}_{\rm J}}}

\def\del#1{{}}

\title[Planet-metallicity correlations]{Tidal Downsizing
  model. II. Planet-metallicity correlations}

\author[S. Nayakshin]{Sergei Nayakshin\\ 
Department of Physics \& Astronomy,
  University of Leicester, Leicester, LE1 7RH, UK\\
{E-mail:~} {\rm Sergei.Nayakshin@le.ac.uk}}

\begin{document}

\date{Received}

\pagerange{\pageref{firstpage}--\pageref{lastpage}} \pubyear{2008}

\maketitle

\label{firstpage}

\begin{abstract}
Core Accretion (CA), the de-facto accepted theory of planet formation,
requires formation of massive solid cores as a prerequisite for assembly of
gas giant planets. The observed metallicity correlations of exoplanets are
puzzling in the context of CA. While gas giant planets are found
preferentially around metal-rich host stars, planets smaller than Neptune
orbit hosts with a wide range of metallicities. We propose an alternative
interpretation of these observations in the framework of a recently developed
planet formation hypothesis called Tidal Downsizing (TD). We perform
population synthesis calculations based on TD, and find that the connection
between the populations of the gas giant and the smaller solid-core dominated
planets is non linear and not even monotonic. While gas giant planets formed
in the simulations in the inner few AU region follow a strong positive
correlation with the host star metallicity, the smaller planets do not. The
simulated population of these smaller planets shows a shallow peak in their
formation efficiency at around the Solar metallicity.  This result is driven
by the fact that at low metallicities the solid core's growth is damped by the
scarcity of metals, whereas at high metallicities the fragments within which
the cores grow contract too quickly, cutting the core's growth time window
short. Finally, simulated giant gas planets do not show a strong host star
metallicity preference at large separations, which may explain why one of the
best known directly imaged gas giant planet systems, HR 8799, is metal poor.
\end{abstract}

\begin{keywords}
\end{keywords}

\section{Introduction}\label{sec:intro}

Given the bewildering diversity of exoplanet system architectures
\citep[e.g.,][]{BatalhaEtal13,WF14}, it is clear that the outcome of planet
formation is a highly stochastic process. Any statistical trends found in the
observed planetary populations have special significance as they testament to
planet-forming processes so robust that they rise above the stochasticity.  A
successful planet formation theory must reproduce such trends.

Metallicity correlations of the observed planets is one such
correlation. Giant planets are detected much more frequently around metal-rich
stars than around metal-poor ones \citep{Gonzalez99,FischerValenti05}. This
correlation has been argued to provide a direct support to Core Accretion
theory for planet formation \citep[e.g.,][]{PollackEtal96} since metal-rich
environments assemble massive cores much more readily
\citep[e.g.,][]{IdaLin04,IdaLin08,MordasiniEtal09a}. These cores are the
crucial step towards forming a gas giant planet in CA framework, thus a
positive giant planet--metallicity correlation ensues. Gravitational disc
Instability model for planet formation \citep[e.g.,][]{Boss98,HelledEtal13a}
was argued to produce a negative rather than positive giant
planet--metallicity correlation since planets' radiative cooling is the
fastest at low dust opacities \citep{HB11}, increasing planet's chances of
survival.

However, there are both observational and theoretical reasons to study the
issue further. Radial velocity observations show that Neptune-mass planet
occurrence does not seem to correlate with the host star's metallicity for FGK
stars \citep{SousaEtal08}. Similarly, transit method observations with {\em
  Kepler}, sensitive to the planet's radius rather than its mass, show
\citep{BuchhaveEtal12} that planets smaller than $4R_\oplus$ form around hosts
with a wide range of metallicities, with the average close to the Solar
metallicity. This is clearly surprising in the context of CA, since these
smaller planets are the precursors of the gas giant planets. If CA's
explanation for the positive gas giant planet correlation with metallicity is
correct then there should be more cores at higher metallicities, which is not
what is observed.

On the theoretical side, several important extensions to the classical variant
of the Gravitational Instability (GI) model for planet formation have been
recently proposed. In particular, GI model that includes planet migration and
pebble accretion was showed to produce a strong positive and not negative
correlation with the star's metallicity \citep{Nayakshin15a,Nayakshin15b} for
{\em coreless} gas giant planets. The goal of this paper is to extend this
recent work on giant planets with cores and also on core-dominated (smaller)
planets and to compare the results with the observations.

The key reason why GI theory may require a major overhaul is the realisation
that GI gas fragments can also migrate from $\sim 100$~AU all the way into the
inner disc \citep{BoleyEtal10}. While planet migration was a standard feature
for the CA model since 1996 \citep{Lin96}, GI planets were somehow thought to
not migrate until recently. If this were true than GI could at best account
for important but rare giant planet systems like HR 8799 imaged directly
\citep[e.g.,][]{MaroisEtal08}, and could never explain the Solar System:
massive self-gravitating proto-planetary discs can hatch clumps only at
$R\simgt $ many tens of AU \citep[e.g.,][]{Rice05,Rafikov05}.

However, simulations, starting with \cite{VB05,VB06}, showed that massive gas
fragments do migrate inward. Furthermore, \cite{BoleyEtal10} suggested a new
way of forming terrestrial-like planets inside $\sim$ a few Jupiter masses gas
clumps by grain growth and sedimentation \citep[see also][]{WC71,Boss97}, and
then releasing them back into the disc by destroying the gas clumps via tidal
forces from the host star. \cite{Nayakshin10c} used analytical estimates of
the relevant processes and arrived at similar ideas, proposing the Tidal
Downsizing (TD) hypothesis for formation of all types of planets observed so
far.

The key question for a planet forming in the framework of the TD hypothesis is
how does the clump's inward migration time scale, which may be as short as
$\sim 10^4$~years \citep{BaruteauEtal11,MichaelEtal11}, compare with the time
scale for its internal evolution? If the fragment contraction time is shorter
than the inward migration time, then the fragment collapses, becoming a bona
fide "hot start" protoplanet, which may mature into a gas giant planet. If the
fragment contraction time is long, then the host's tidal forces catch up with
the fragment when the latter migrates too close to the star, and the fragment
is disrupted.  The fragment's short existence may however not be all in vain
as there could be a remnant. If grains within the fragment had a sufficient
time to grow and sediment to the centre, getting locked into a
self-gravitating massive core \citep[see
  also][]{Kuiper51,McCreaWilliams65,WC71,Boss97,HS08}, then the remnant of the
disruption is the core -- a practically ready rocky planet. No planetesimal
accretion is required for the planet to survive at that stage, although
"veneer accretion" of them or pebbles \citep{JohansenLacerda10} is possible.


Making detailed predictions in the TD hypothesis setting worthy of comparison to modern
observations of exoplanets is however not trivial. Until recently, it was
assumed that collapse of the self-gravitating GI gas fragments must occur due
to radiative cooling
\citep[e.g.,][]{Bodenheimer74,BodenheimerEtal80,CameronEtal82,HelledEtal08,Nayakshin10b,NayakshinCha13}. The
formation of giant gas planets in this picture is similar to that of low mass
stars \citep[e.g.,][]{Larson69}, except that accretion of gas onto the
pre-collapse planets is terminated very early on
\citep[e.g.,][]{Nayakshin10a}. This formation channel favors survival of
massive ($M_p \simgt$ a few $\mj$) gas giants \citep{ForganRice13b}, as more
massive planets cool more rapidly \citep[e.g., see][]{Nayakshin15a}. Also, radiative collapse of H$_2$-dominated fragments produces a negative planet frequency --
metallicity correlation \citep{HB11} because radiative cooling is more rapid
at low metallicities.  That is clearly at odds with observations
\citep{Gonzalez99,FischerValenti05}.

We \citep{Nayakshin15a}, however, considered accretion of $\simlt 1$ cm sized
grains \citep[usually called "pebbles";][]{JohansenLacerda10,OrmelKlahr10}
onto the pre-collapse fragments. While gas has pressure gradient forces that
may prevent its accretion from the protoplanetary disc onto the fragments
\citep[e.g.,][]{NayakshinCha13}, pebbles weakly coupled to the gas by
aerodynamical forces can still accrete onto the fragment in an analogy to the
pebble accretion process onto massive solid cores \citep{LambrechtsJ12} in the
context of the CA model.

Surprisingly, pebble accretion on GI fragments was found to accelerate their
contraction despite increasing their metallicity and hence
opacity. Physically, addition of pebbles to the fragment increases its weight
without increasing its thermal energy\footnote{This is in difference to
  accretion of planetesimals in the CA framework, in which planetesimals
  impact the growing planet at a high (up to a few km s$^{-1}$) velocity and
  hence may actually heat it. In contrast, pebbles sediment onto the fragment
  at a moderate $\sim $ a few m/sec velocity, or else they fragment. Their
  kinetic energy input into the planet is hence negligible.}. Since
pre-collapse molecular gas fragments are polytropes with an effective
polytropic index $\gamma$ edging ever closer to the unstable value of $4/3$ as
the central temperature of the fragment increases towards 2000~K
\citep[see][]{Nayakshin15a}, addition of a relatively small, $\sim 10$\%,
amount of mass in metals to the fragment usually tips it over into
collapse. Higher metallicity environments provide larger pebble accretion
rates, hence \cite{Nayakshin15b} found, via a population synthesis like study
of {\em coreless} gas fragments, that the frequency of gas giants survived the
initial tidal disruption increases as metallicity of the host increases.

An important issue not considered by \cite{Nayakshin15a,Nayakshin15b} is
formation of a core within the fragment. If pebbles entering the fragment
accrete onto the core, the core's accretion luminosity may be significant and
this may lead to expansion of the fragment instead of its contraction. A
similar effect -- accretion of planetesimals onto the growing core -- is well
known in the context of the CA model and has to be taken into account when
determining the fate of the planet
\citep[e.g.,][]{PollackEtal96,AlibertEtal05}. \citet[paper I
  hereafter]{Nayakshin15c} has recently presented numerical algorithms that
extend the work of \cite{Nayakshin15b} by adding the core formation and growth
inside the pre-collapse fragments.

Here we use that framework to study the planet frequency of survival versus
metallicity dependence for all types of planets rather than just the coreless
giants. We also go beyond of our previous work by studying not just the
outcome of the migration vs disruption competition, but also how the
population of planets formed in TD model would look like in the planet mass --
separation plane.

The structure of our paper is as following. In \S \ref{sec:numerics} we
present initial conditions, assumptions and numerical algorithms that are
employ to calculate an outcome of a gas-dust fragment being born in an outer
region of a massive protoplanetary disc. In \S \ref{sec:examples}, two example
planet formation tracks are shown, one for a gas giant and another for a hot
Super-Earth planet. In \S \ref{sec:grid} we present assumptions and methods
for calculating a grid of models in a population synthesis like study.  \S
\ref{sec:results} presents the population synthesis study result in the planet
mass versus planet-host separation plane. In \S \ref{sec:zcorr} we marginalise
over the results to derive the planet-metallicity correlations for our
simulated planets. Discussion of main ideas and results of the paper is given
in \S \ref{sec:discussion}, whereas conclusions are presented in \S
\ref{sec:conclusions}.

\section{Numerical methods}\label{sec:numerics}

\subsection{Disc treatment}\label{sec:disc}

We follow the procedures described in paper I with several changes detailed
now. Our goal is to simulate formation of planets from their birth to the end
of the disc migration phase. The initial disc surface density profile is given
by
\begin{equation}
\Sigma_0(R) = {A_m \over R} \left(1 - \sqrt{R_{\rm in}\over
  R}\;\right)\;\exp\left[-{R\over R_0}\right]\;,
\label{sigma_init}
\end{equation}
where $R_0$ is the disc length-scale, set to $R_0 = 100$ AU, $R_{\rm in} =
0.08$ AU is the inner boundary radius. The constant $A_m$ is calculated so
that the disc contains a given initial mass $M_{\rm d} = 2\pi \int_{R_{\rm
    in}}^{R_{\rm out}} R dR \Sigma_0(R)$. The disc is "live", that is evolved
by solving the time-dependent 1D viscous disc evolution equations that include
the planet-disc interactions (section 3 in paper I), and now also include the
disc photo-evaporation as a sum of the UV and the X-ray driven terms, using
the fits to the photo-evaporation rates $\dot\Sigma_{\rm ev}(R)$ from
\cite{AlAr07} and \cite{OwenEtal12}, respectively. The ionising photon
luminosity of the star is set to $\Phi_{\rm ion} = 10^{42}$ photons s$^{-1}$,
while the X-ray flux of the star is $L_X = 2\times 10^{30}$ erg~s$^{-1}$.

\begin{figure}
\centerline{\psfig{file=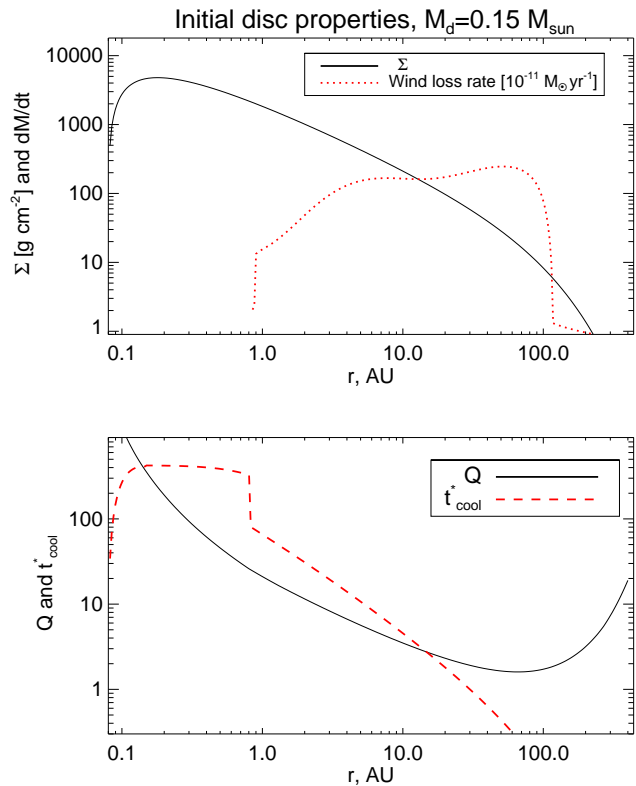,width=0.5\textwidth,angle=0}}
\caption{Initial disc properties for $M_d = 0.15 \msun$. Top: $\Sigma(R)$ and
  the wind mass loss rate profile, defined as $\dot\Sigma(R)\pi R^2$, as
  labelled. Bottom: Toomre parameter, $Q$, and dimensionless cooling time of
  the disc, $t_{\rm cool} \Omega(R)$, versus radius. The disc is
  gravitationally unstable and may fragment at $R\sim 70-80$~AU.  }
\label{fig:disc_ics}
\end{figure}

Gas fragments in the disc are born by self-gravitational instabilities, hence
the disc must be initially gravitationally unstable in its outer
region. Figure \ref{fig:disc_ics} shows the initial disc properties for disc
mass $M_d = 0.15 \msun$ and stellar mass $M_*=1\msun$. The top panel shows the
disc surface density, $\Sigma(R)$, as well as the disc photo-evaporation rate
profile (which does not change during the simulation). The quantity plotted
with the red dotted curve is $\dot\Sigma_{\rm ev}\pi R^2$, in units of
$10^{-11} \msun$ year$^{-1}$. The bottom panel shows the Toomre $Q$-parameter
(black solid curve) and the dimensionless cooling time, $t^*_{\rm cool} =
t_{\rm cool} \Omega(R)$. For the disc to be self-gravitating and actually
fragmenting, $Q\simlt 1.5$ and $t^*_{\rm cool} \simlt 3$
\citep{Gammie01}. Both of these conditions are satisfied at $R\sim 70-80$~AU.

The disc viscosity is described with the \cite{Shakura73} viscosity parameter,
but we now add a self-gravity contribution to it:
\begin{equation}
\alpha = \alpha_0 + \alpha_{\rm sg}\;,
\label{alpha0}
\end{equation}
 where $\alpha_0$ is radius and time-independent (free) parameter of the
 model, whereas $\alpha_{\rm sg}$ describes gravito-turbulence
\begin{equation}
\alpha_{\rm sg} = 0.2 {Q_0^2 \over Q_0^2 + Q^2}\;.
\label{alpha_sg}
\end{equation}
where $Q_0=2$ and $Q$ is the local Toomre's parameter. This form is inspired
by \cite{Lin87} suggestion that $\alpha\propto Q^{-2}$ in the self-gravitating
regime, on the one hand, and more recent simulations showing that $\alpha_{\rm
  sg}$ saturates at about 0.1 \citep{Rice05}, on the other.

\subsection{Planet's birth and migration}\label{sec:migration}

The following considerations are important for both the initial and the final
phases of our simulations. Simulations of self-gravitating protoplanetary
discs show that fragments rarely form in isolation, hence we believe that
every star is likely to have hatched a number of fragments early
on. Presumably, most of these fragments migrate rapidly and perish by being
disrupted \citep[see, in particular,][]{Vorobyov13c} or by being driven all
the way into the star. This picture is consistent with the suggestion that
disruption and swallowing of planets by their host stars power FU Ori
outbursts of young protostars \citep{VB06,BoleyEtal10,NayakshinLodato12},
statistics of which suggests that every star may have $\sim 10-20$ of such
episodes \citep{HK96}.

Given that the observed giant gas planets' frequency of occurrence is $\simlt
10$\% \citep{WF14} per star, the survival of a giant gas planet is a
very rare event indeed. It is reasonable to assume that giant gas planets that
survive to be observed are those that were either hatched by the disc late, or
those that hang around at the outer disc for a while before migrating in,
because such fragments are more likely to avoid the tidal disruption and also
being driven all the way into the star. 

This view is not unreasonable. Simulations show that presence of one clump in
a disc may induce formation of more clumps in the same disc
\citep[e.g.,][]{Meru13}. Multiple clump formation leads to strong interactions
between the clumps, so that some of them migrate inward faster while others
are scattered on larger orbits \citep[e.g.,][]{ChaNayakshin11a}; some of the
clumps may be completely ejected from the system by the clump-clump
interactions \citep{BV12}. Clumps on scattered but still bound orbits may take a while to
loose their orbital inclinations with respect to the midplane of the disc,
and eccentricities, and only then start migrating in.

Furthermore, presence of clumps in the disc may in fact change the
fragmentation properties of the disc since the clump induces additional
perturbations \citep[see again][, where discs were found to fragment at
  surprisingly small radii in the presence of a pre-existent gas clump at
  larger radii]{Meru13}. \cite{NayakshinCha13} focused on migration of a
single fragment in a marginally self-gravitationally stable disc, thus placing
the fragment in a massive but {\em stable} disc that would not fragment on its
own. They found that in the presence of the clump the disc became more
unstable and formed additional fragments. This shows that it is possible to
hatch new gas clumps in discs with $Q > 1.5$ if there are already fragments
born earlier. 

Clearly, when survival of the fragments is rare, we should not under-estimate
the importance of initial conditions giving the clumps the best chance of
surviving. Therefore, our population synthesis model aims to account for this
by (a) slowing down the type I migration below its nominal rate by some factor
larger than unity (see below); (b) allowing the disc mass to be below the
$M_d=0.15 \msun$ marginally gravitationally unstable configuration when the
fragment is born, and (c) removing the disc instantaneously after a time
$t_{\rm rem}$ randomly sampled between 0.5 and 5 Million years. These choices
are certainly not unique but we feel they are reasonable. An extremely rapid
external photo-evaporation of the disc by a close-by massive star
\citep{Clarke07} or the presence of a secondary at beyond 100 AU that is
later removed by star-star interactions may be important for such a rare
population of planets as the gas giants.

The disc exchanges angular momentum with the planet via type I and/or type II
migration torques. The only difference from paper I is that the type I
migration time scale, $t_{I}$, is now given by
\begin{equation}
t_{I} = f_{\rm mig} {M_*^2 \over M_p M_{d}} {a^2\over H^2} \Omega_a^{-1}\; \;.
\label{time1}
\end{equation}
The factor $(1+M_p/M_d)$, used in paper I on the right hand side of this
equation, has been dropped, as it is argued that the saturation of the type I
migration rate in the limit $M_p/M_d \gg 1$ is automatically taken into
account by passing the torque of the planet to the 1D viscous disc evolution
code. We find that when this torque is large then a gap may start opening,
thus reducing the torque.

In equation (\ref{time1}) $f_{\rm migr} > 1$ is a free parameter that accounts
for a slower migration rate of the planet in a non self-gravitating disc
compared to the isothermal type I migration rate. Since our discs are sampling
the post self-gravitating phase of the disc evolution by design (see below),
we may expect type I migration to be slower than that for $Q\sim 1$ discs
\citep{BaruteauEtal11}, for which $f_{\rm migr}\sim 1$. In the models below
$f_{\rm migr}$ is randomly sampled between $1$ and 10.

\subsection{Planet's internal evolution}

Planet's internal evolution is calculated as in paper I, utilising the "follow
the adiabats" approximation to convective/radiative cooling of the planet. The
planet is initialised as a polytropic sphere of gas of homogeneous
composition, with metallicity, $z_0$, equal to that of the parent disc, with a
given central temperature $T_0$ (see \S \ref{sec:grid}).  The planet cools by
emission of radiation assuming interstellar dust opacity \citep{ZhuEtal09}
multiplied by the factor $f_{\rm op}<1$ to account for grain growth and also
by the metallicity of the planet in Solar units, that is, by
$z/z_\odot$. Since the planet is surrounded by the disc, its irradiation by
the disc (or by the star if the planet is directly exposed to it in the final
stages of calculation) reduces the rate at which the planet cools, creating a
thermal bath effect \citep{CameronEtal82,VazanHelled12}.

The planet accretes pebbles from the surrounding disc at the Hill's accretion
rate \citep{LambrechtsJ12}
\begin{equation}
\dot M_z = 2 f_{\rm p} \Sigma_g(a) \Omega_a R_H^2
\label{dotmz}
\end{equation}
where $R_H$ is the planet's Hills radius, $\Omega_a = (GM_*/a^3)^{1/2}$, and
$\Sigma_g = f_g \Sigma(a)$ is the grain surface density at radius $R=a$. The
pebble mass fraction, $0< f_{\rm p}< 1$, is a free parameter independent of the
planet's location, fixed for a given run (see \S \ref{sec:grid} below).

Note that pebble accretion rate depends directly on the surface density of gas
around the planet. If the planet manages to open a deep gas in the disc,
$\Sigma(a)$ drops to very low values and hence $\dot M_z$ dives to negligible
levels. Pebble accretion is also turned off if and when the fragment undergoes
the second (Hydrogen molecule dissociation) collapse. It is argued that after
the collapse the planet is so compact that it should be able to accrete gas
from the disc as well, so that both gas and pebbles would be accreted. We
currently do not include gas accretion onto post collapse planets.

Pebbles entering the fragment have a fixed initial size of $a_{\rm peb} =0.03$
cm for this paper, and are deposited in the outer several mass grids of the
planet, where they are mixed homogeneously with the grains already in those
regions. The grains are allowed to grow by sticking collisions and sediment
into the centre of the fragment into the core, as described in paper
I. Turbulence within the fragment, parameterised by a coefficient $\alpha_d$,
and convection, however, tend to counteract grain sedimentation. We
usually find that grain size must increase to about $a_g \sim 1$ cm before
they sediment efficiently. A further significant obstacle to a rapid grain
sedimentation into the core is the fact that grains fragment in collisions
rather than stick if collision velocity is too high, as shown by laboratory
experiments on dust growth
\citep[e.g.,][]{BlumMunch93,BlumWurm08,BeitzEtal11}. As explained in paper I, section
4.3, grain growth turns into grain fragmentation if grain sedimentation
velocity with respect to the surrounding gas exceed $u_{\rm max}$, set in this
paper to 3 m~s$^{-1}$. Grains are also vaporised if the surrouding gas
temperature is too high for the given grain species.

Three grain species are included in the model: water ice, CHON, and silicates
("rocks"). The relative abundances of the species are 0.5, 0.25 and 0.25,
respectively. The grain growth, fragmentation, vaporisation, and sedimentation
equations are solved for each species separately since these three species
have very different material and thermal properties.

Grains reaching the centre are allowed to accrete onto the solid core there,
whose initial mass is set to a very "small" value ($10^{-4} \mearth$). Growing
core is expected to radiate some of its gravitational potential energy away,
but a self-concistent modelling of energy transfer within the core is fraught
with many physical uncertainties \citep[e.g.,][]{StamenovicEtal12}. As in
paper I (section 4.4), we parameterise the energy release by the core via the
Kelvin-Helmholtz contraction time, $t_{\rm kh}$, of the solid core, which is
fixed for all the runs here at $t_{\rm min}=3\times 10^5$~years. The
luminosity released by the core is an internal energy source for the
contracting fragment, analogous to that of a stellar core, and may even exceed
the luminosity of the whole planet, causing its expansion.

\section{Example runs}\label{sec:examples}

Before the grid of models is discussed, we present two example planet
formation tracks that illustrate the sequence of events in the TD model for planet
formation.

\begin{figure}
\centerline{\psfig{file=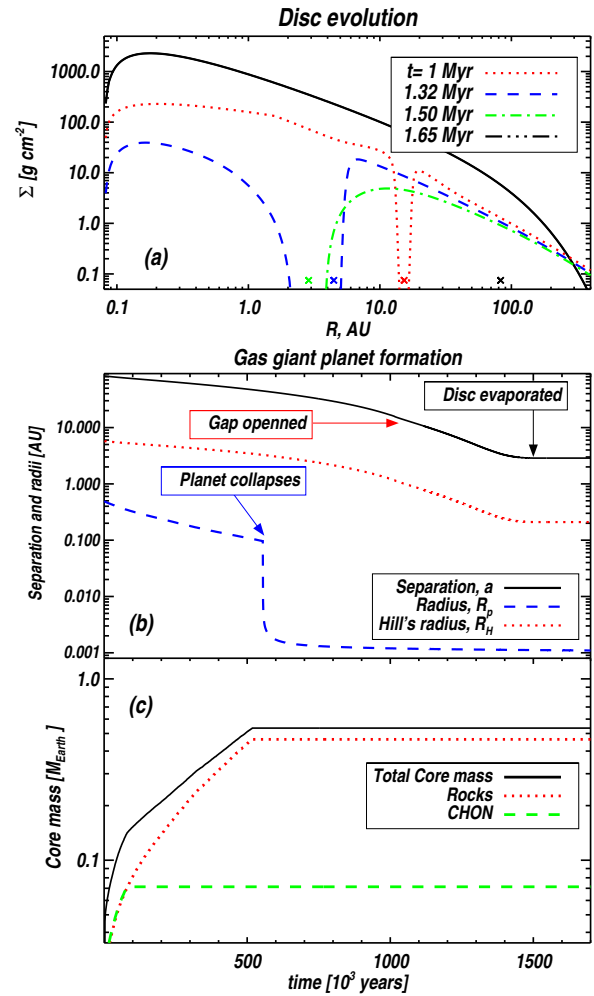,width=0.5\textwidth,angle=0}}
\caption{Evolution of the disc (panel a) and the embedded fragment (panles b
  and c) that survives to be a gas giant planet (see \S \ref{sec:giant} for
  detail). Panel (a) shows disc surface density profiles at times $t=0$ (solid
  curve) plus several later times as labelled in the legend. The position of
  the planet at corresponding times is marked by a cross of same collor at the
  bottom of the panel. Panel (b) shows the planet's separation, radius and the
  Hills radius, whereas panel (c) shows the mass of the core versus time.}
\label{fig:disc_planet_giant}
\end{figure}

\subsection{A gas giant planet}\label{sec:giant}

Figure \ref{fig:disc_planet_giant} shows an example selected from the grid of runs in
which the end result is a gas giant planet of mass $M_p= 1.17 \mj$ at
separation of $a=2.7$ AU. The initial mass of the planet is $M_p= 1.0 \mj$, so
all the extra mass in the end of the run comes from accretion of pebbles from
the disc. The initial mass of the disc is $M_{\rm d} = 0.074\msun$, planet
location $a_0=83$ AU, and the initial planet's central temperature $T_c =
254$~K. The disc metallicity is Solar, that is, $z=0.015$, the pebble mass
fraction is $f_{\rm p} = 0.086$, disc viscosity parameter $\alpha_0 =
2.9\times 10^{-3}$. The disc removal time is $t_{\rm rem} = 3$ Million years,
but in this simulation the disc dissipates earlier due to photo-evaporation,
as we shall see below. Further, the opacity reduction factor $f_{\rm op} =
0.5$, the planet's type I migration speed is moderated by factor $f_{\rm migr}
= 8.05$, which is on a high side for the grid of models, so this run presents
an example of a rather slowly migrating clump.

Fig. \ref{fig:disc_planet_giant}a presents the disc surface density profile at
five different times. The initial condition at $t=0$ is shown with the black
solid curve, all the other curves are marked by their respective times in the
legend. The crosses on the bottom of the panel show the position of the planet
at times colour-coded in the same way as the surface density curves.

Next panel, fig. \ref{fig:disc_planet_giant}b, reports the radial location of
the planet (black solid curve), the planet's Hill (red dotted) and its actual
($R_p$, blue dashed) radii, all ploted versus time. Boxed text and the
corresponding arrows mark the times of three important transitions in this
run. In particular, the fragment is initially molecular Hydrogen dominated,
collapses at time $t=0.55$ Million years, at which point Hydrogen becomes
atomic and partially ionized. During collapse the fragment contracts to the
radius of just a few times that of Jupiter, and then continues to contract
further due to a rapid radiative cooling. This fragment is never challenged by
tidal forces of the star as the planet's Hill radius (red dotted curve) is
always much larger than $R_p$.

The next notable transition in the planet-disc system occurs at $t\approx 1$
Million years, when the planet opens a deep gap in the disc. The planet is
located at $a\approx 27$~AU at that time. From that time on, the planet
migrates in type II regime. Note that at time $t=1.32$~Million years (the
dashed blue curve in fig. \ref{fig:disc_planet_giant}a), the gap becomes
deeper than it was before, and the inner disc has a significant gas deficit at
$a \sim 2$~AU. This is not only due to the gap openning; the photo-evaporation
of the disc is especially strong at $a\sim$ a few AU, and is negligible within
1 AU, as can be seen from fig. 1. The planet-facing side of the disc inside
the planet's orbit thus loose mass rapidly due to photo-evaporation.

At this point in time, the inner disc is drained roughly equally efficiently
by both accretion onto the star and photo-evaporation. By time $t=1.5$ Million
years, there is a complete gap between the planet and the star. The outer disc
is then evaporated too, and the planet stops migrating completely.

Finally, fig. \ref{fig:disc_planet_giant}c shows the core mass assembled
within the fragment, and its decomposition on the silicates (``rocks") plus
water ice and CHON. Since the fragment heats up quickly when pebble accretion
commences, water ice grains vaporise before they could sediment into the core,
hence they do not contribute to the core at all. CHON grains (green dashed
curve) sediment while the centre of the fragment is cooler than $T\approx 400$
K, and then only the rocks are refractory enough to continue to accrete onto
the core. This particular planet did not manage to assemble a massive solid
core before it collapses (when all grains not yet locked into the core are
vaporised); the final core mass is only $M_{\rm c} = 0.55\mearth$.

The final metallicity of the planet is $z=0.158$, about ten times that of the
initial (Solar) disc metallicity. This example thus shows how a metal-rich gas
giant planet with a low mass core can be assembled in the context of the TD
hypothesis.

\subsection{A hot Super-Earth planet}\label{sec:hse}

\begin{figure}
\centerline{\psfig{file=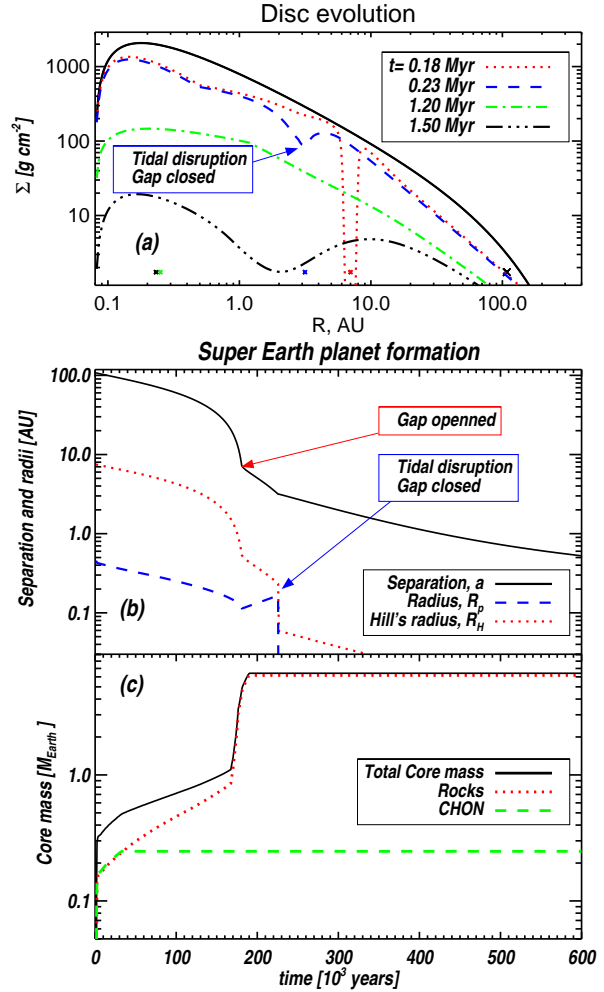,width=0.5\textwidth,angle=0}}
\caption{Same as Fig. \ref{fig:disc_planet_giant}, but for a simulation that
  forms a hot Super-Earth planet as the end result (\S \ref{sec:hse}). The
  blue box in the top panel (a) shows where and when the gas fragment is
  tidally disrupted, which leaves behind a massive $M_{\rm c}\sim 6.4\mearth$ core
  that continues to migrate, and finally arrives in the ``hot region'',
  $a=0.23$, by the time the disc is dispersed.}
\label{fig:disc_planet_SE}
\end{figure}

Figure \ref{fig:disc_planet_SE} demonstrates how a hot Super-Earth planet can
be assembled in the framework of TD.  The
initial mass of the planet is again $M_p= 1.0 \mj$. The initial mass of the
disc is $M_{\rm d} = 0.066\msun$, planet's location is $a_0=108$ AU, and the
initial planet's central temperature $T_c = 282$~K. The disc metallicity is
$z=1.7$ times Solar, the pebble mass fraction is $f_{\rm p} = 0.08$, disc
viscosity parameter $\alpha_0 = 2.3\times 10^{-3}$. The disc removal time is
$t_{\rm rem} \approx 5$ Million years, but again disc dissipates earlier due
to photo-evaporation. The planet's type I migration speed is barely reduced
from the isothermal result, e.g., factor $f_{\rm migr} = 1.3$, so the fragment
migrates in much sooner than it does in \S \ref{sec:giant}. The maximum grain
sedimentation velocity for this simulation is set to $v_{\rm max} =
10$~m~s$^{-1}$, which, combined with a higher $z$ value, encourages a faster
solid core growth.

It appears that the much more rapid migration is to blame for most of the
differences in the fate of the fragment here as compared to the much more
slowly migrating one in \S \ref{sec:giant}. Fig. 3 shows that the fragment
finds itself in the inner disc much quicker than the one in Fig. 2 did. The
gap in the disc is openned by the fragment at around $a = 7.2$~AU, at time $t
= 0.18$ Million years. Significantly, the fragment is still in the
pre-collapsestate at that moment, and it also has a massive solid core,
$M_{\rm c} = 4.8 \mearth$ at the time of the gap openning. The more massive
core is due to the higher value $v_{\rm max}$ in this simulation.

Contraction of the fragment in this simulation occurs almost entirely due to
pebble accretion because radiative cooling is very inefficient. This is
because the fragment's dust opacity is high, especially so due to pebble
accretion: by $t=0.18$ Million years the fragment metallicity is $z=
0.144$. The fragment is also basked in the radiation field of the disc (which
at the outer disc radii is dominated by irradiation from the star). Now, when
pebble accretion onto the fragment stops, its contraction stops as
well. However, since the fragment is metal-rich, the {\em core of the
  fragment} continues to grow in mass rapidly and its luminosity becomes
sufficiently high to start puffing the fragment up. This increase in the
fragment's radis after the disc gap openning is easy to spot in Fig. 3b.
Since the fragment continues to migrate in, now in type II regime, the
planet's Hills radius continues to shrink, whilst the planet's radius, $R_p$,
increases with time. A tidal disruption is thus unavoidable, and indeed occurs
when the fragment reaches $a=3.16$~AU.

The core's mass is then equal to $M_{\rm c} = 6.4 \mearth$. As the fragment is
disrupted, its mass plumets to $M_{\rm c}$ (in this paper we neglect a
possible tightly bound dense gas ``atmosphere'' around the core, which is not
likely to be massive as our cores are quite bright). With this reduced mass,
the planet is no longer able to maintain the deep gap in the disc, which is
swiftly closed (see the blue dashed curve in Fig. \ref{fig:disc_planet_SE}a).
The mass of the core is however sufficiently high for it to continue its
migration in type I now, on a longer time scale yet fast enough to arrive
at $a= 0.23$~AU by the time the disc is dispersed at $t=1.5$
Million years.

The composition of the core is almost exclusively rock, with CHON and
especially water ice grains unable to sediment because the core heats up to
temperatures above $\sim 400$ K quickly (cf. paper I for more detail). The
core's composition is significantly different from that predicted by the Core
Accretion paradigm, in which ices are very important
\citep{PollackEtal96,AlibertEtal05} for assembly of cores as massive as this
one. This may be a testable difference between the models if composition of
Super-Earths is constrained observationally, although the potential presence
of a Hydrogen/He atmosphere on top of the core may complicate differentiation
between the models.

\section{The population synthesis grid of models}\label{sec:grid}

A population synthesis involves investigation of the model's parameter space,
which is usually large, by randomly sampling it
\citep[e.g.,][]{MordasiniEtal09a} in order to pull out statistical trends of
the model that may not be distinguisheable from just a handful of runs. In
following this approach, we fix some parameters of the model in order to
reduce the parameter space to a manageable but physically meaningful minimum.

Table 1 lists these fixed parameters, their meanings and values. The latter
are chosen to be ``reasonable'' based on experiment (e.g., $v_{\rm max}$) or
on what is commongly used in literature (e.g., $L_X$). While these values are
not unique, we found by varying them that none of the parameters in Table 1
change the metallicity {\em trends} that are the focus of this paper
significantly, although quantitatively there are of course changes in the
results. For example, the normalisation of the giant planet number per star
surviving in the end of the simulations does depend on how the disc is removed
in the end of the simulations, and hence on $L_X$ and $\Phi_{\rm ion}$, but
the metallicity trends of the survived planets are the same. 

One parameter in Table 1 that is difficult to constrain from first principles
and which does influences the results is $f_{\rm op}$, the grain opacity
reduction due to grain growth. The latter is fixed here on a rather large
value, $f_{\rm op}=0.5$, which essentially precludes a significant opacity
reduction. This is done following the arguments presented in paper I: the
observed metallicity trends are incompatible with low dust opacity models and
presumably imply that dust fragmentation (and not only dust growth) is
occuring within the fragments to keep the opacity at relatively large values.

Table 2 lists the set of variable parameters, and their minimum and maximum
values. The parameters for a run are then randomly picked in the logarithmic
space between these respective minimum and maximum values. For example, the
disc viscosity parameter, $\alpha_0$, is very poorly known for protoplanetary
discs (and any other astrophysical discs). We introduce a random variable
$\xi_\alpha$ uniformly distributed between $0$ and 1, draw a random value
$\xi_\alpha$, and define $\alpha_0$ for a given simulation by
\begin{equation}
\ln \alpha_0 = \xi_\alpha \ln \alpha_{\rm min} + \left( 1 - \xi_\alpha\right)
\ln \alpha_{\rm max}\;.
\label{alpha_r}
\end{equation}
Similarly, random uniformly distributed variables are generated for all the
other entries in Table 2.

Having defined the parameters for a simulation, we model one gas fragment per
disc at a time. The fragment starts with a given initial total (H/He + grains)
mass, and has the same metallicity as that of the disc. The initial
temperature is randomly picked (as described in equation \ref{alpha_r} above),
between $T_0(M_p)$, given by
\begin{equation}
T_0(M_p) = 100 \hbox{ K } \left[{M_p \over 0.5 \mj}\right]^{1/2}\;,
\label{T0}
\end{equation}
and twice $T_0(M_p)$. Here the mass dependence of $T_0$ reflects the fact that
more massive fragments cool more rapidly
\citep[e.g.,][]{HelledEtal08,Nayakshin10a}, and are also hotter at formation
due to the adiabatic compression. The choice made in equation \ref{T0} does
not pre-determines or drives the main results of this paper.

Finally, we wish to investigate how our results depend on the metallicity of
the host star/disc and the initial mass of the fragment. Therefore, the random
models are repeated for a grid of the metallicity and the initial fragment's
mass values. The metallicity grid is given by $z_i/z_\odot = 3^{(i-3)/2}$,
with $i = 1, 2$, ... 5. The initial planet mass grid is given by $M_p/\mj =
0.5$, 0.7, 1 and 2. For each of the combinations of $z$ and $M_p$ from this
grid, e.g., $z = 3 z_\odot$ and $M_p = 1\mj$, the total of 243 models are
run. This gives the total of 4860 runs. A typical run takes about 2 hours of
physical time on a single CPU, although some runs (usually with higher disc
mass and viscosity, as that leads to smaller time steps) had to be executed
for up to 48 hours.

\begin{table}
\caption{Fixed parameters of the simulations and their values: maximum
  velocity for sticking rather than fragmentation, $v_{\rm max}$, in
  m~s$^{-1}$; core's Kelvin-Helmholtz contraction time, $t_{\rm kh}$, in years;
  opacity reduction factor due to dust growth, $f_{\rm op}$; pebble radius,
  $a_{\rm peb}$, in cm; ionising photon flux from the star, $\Phi_{\rm ion}$,
  photons s$^{-1}$; X-ray luminosity of the star, $L_X$, in erg s$^{-1}$.  }
\begin{tabular}{lcccccc}
\hline
Parameter & $v_{\rm max}$ & $t_{\rm kh}$ & $f_{\rm op}$ & $a_{\rm peb}$ &
$\Phi_{\rm ion}$ & $L_X$\\ \hline
Value &  3 & $3\times 10^5$  & 0.5 & 0.03 & $10^{42}$ & $2\times 10^{32}$\\ \hline
\end{tabular}
\label{tab:1}
\end{table}

\begin{table}
\caption{Randomly varied parameters of the grid of models. The first row gives
  parameter names, the next two their minimum and maximum values. The columns
  are: $\alpha_0$, the disc viscosity parameter; $f_{\rm p}$, the pebble mass
  fraction; factor; $a_0$ [AU], the initial position of the fragment; $M_{\rm
    d}$, the initial mass of the disc, in units of $\msun$; $f_{\rm migr}$,
  the reduction multiplier for type I planet migration; $t_{\rm rem}$, Million
  years, the time for disc removal; $\alpha_d$, turbulence parameter within
  the fragment.}
\begin{tabular}{lccccccc}
Parameter & $\alpha$ &   $f_{\rm p}$ & $a_0$  & $M_{\rm d}$ &
$f_{\rm migr}$ & $t_{\rm rem}$ & $\alpha_{d}$ \\ \hline
Min &  0.001 &  0.05 &  60  &  0.05 & 1  &  0.5 & $10^{-4}$\\
Max &  0.01  &  0.2  &  120 &  0.2  & 10 &  5   & 0.003 \\ \hline
\end{tabular}
\label{tab:2}
\end{table}

\section{Results}\label{sec:results}

\begin{figure*}
\centerline{\psfig{file=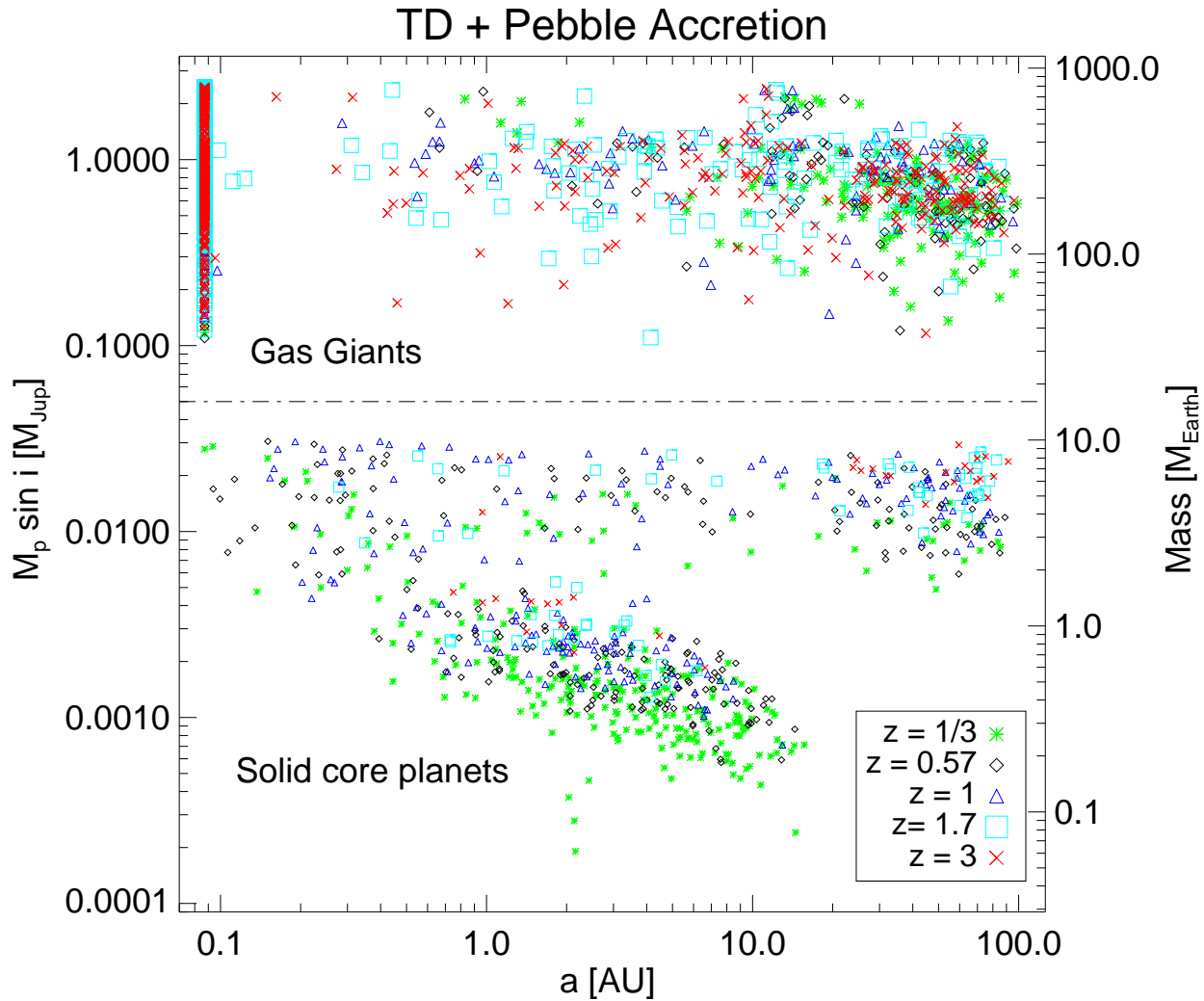,width=0.95\textwidth,angle=0}}
\caption{Planet mass versus planet-host separation for the pebble accretion
  grid of models for different metallicities, as shown in the legend in Solar
  metallicity units. For the gas giant planets (historically found by RV
  methods) $M_p\sin i$ is used as proxy for $M_p$, where $\sin i$ is randomly
  generated orbital inclination with respect to the line of sight. For
  core-dominated planets, located below the dash-dotted line, $M_p$ in units
  of Earth mass (on the right vertical axis) is plotted.}
\label{fig:scatter}
\end{figure*}

Figure \ref{fig:scatter} presents the results of the 4860 runs in the planet
mass versus planet-star separation plane. In order to facilitate a visual
comparison to the exoplanetary data, giant planet masses are shown as $M_p
\sin i$, where $i$ is inclination of the system to the observer, generated
randomly assuming a uniform distribution for $\cos i$. This is done because
many of the known giant planets were detected by the radial velocity method
and so also have the unknown $\sin i$ factor. The gas giants are defined here
as fragments that did not have a tidal disruption and hence are all more
massive than our minimum starting fragment mass, e.g., $0.5 \mj$. All of these
planets are above the horizontal dashed line.

Cores, the remnants of tidal disruption of the initial gas fragments, are the
planets below the dashed line. In Fig. \ref{fig:scatter}, all of the survived
giant planets are shown, but, for the sake of clarity, only a third of the
cores, randomly selected from the full sample, is plotted. The mass of the
cores, the scale for which is given in units of $\mearth$ on the right
vertical axis, is not multiplied by the $\sin i$ factor since many of the
observed planets of this mass are detected by the transit method which is not
affected by the inclination uncertainty.

We emphasise that in this paper no consideration for the presence of a bound
atmosphere around the cores after the disruption is made, so their mass may be
somewhat under-estimated. Further, gas accretion from the surrounding
protoplanetary disc onto the cores and onto the post-collapse giant gas
planets may well occur but is also not taken into account, thus again possibly
under-estimating the final mass of the planets. Finally, an additional
uncertainty which may bring down the mass of the giant gas planets is that
here we assumed that H$_2$ dissociation collapse of a gas giant leads to the
instantaneous collapse of the whole fragment. However, 3D hydrodynamical
simulations of embedded clumps in protoplanetary discs show that only a
fraction of the total mass of the fragment collapses into the post-collapse
planet immediately, with the rest \citep[up to $\sim 50$\% of the fragment's
  mass][]{GalvagniEtal12} collapsing into a circum-planetary disc. It is not
clear whether all of this disc or only a fraction is eventually accreted onto
the planet. There is thus an uncertainty of a factor of $\sim 2$ in the mass
of the giant planets shown in Fig. \ref{fig:scatter}.

Despite these uncertainties, there are clear and strong metallicity trends in
the planet populations in Fig. \ref{fig:scatter} which we shall investigate in
detail shortly. To aid this, the symbols in Fig. \ref{fig:scatter} are
colour-coded to show the metallicity of the host star/disc. For example, the
red colour shows the most metal rich systems, $z=3 z_\odot$, whereas the green
show the most metal-poor systems in the sample, $z=(1/3) z_\odot$. It is easy
to notice that most of the giant planets in the inner few AU of the diagram
are metal rich; this is not so for cores, however. This observation
essentially summarises the main findings of this paper.

Finally, note the vertical column of giant planets at $a\approx
0.09$~AU. These are the fragments that managed to collapse before being
tidally disrupted but that continued to migrate inward rapidly and reached the
inner boundary of our computational domain (we stop the calculation if the
planet reaches a short distance away from $R_{\rm in} = 0.08$ AU). This column
contains 2391 planet, e.g., almost a half of all fragments born in the disc,
managed to collapse before being tidally disrupted but reached the innermost
disc radius. Most of these would probably be completely devoured by the star
if our computational domain were extended all the way to the inner boundary. 

An additional point to make on this is that our calculation is optimised for
survival of the giant planets because we essentially simulate {\it the end}
phase of the disc evolution. There is no reason why fragments could not be
born before the phase we are simulating. These would migrate even faster, and
most likely would all end up either being tidally disrupted or being driven
all the way to the inner boundary. Hence the number of giant planets migrating
all the way into the star is still an under-estimate. This is consistent with
the view presented earlier -- that destroyed gas giant planets cause FU Ori
outbursts and that statistically speaking, there may be many such events per
protostar.

\section{Metallicity correlations}\label{sec:zcorr}

\subsection{Small separations}\label{sec:smallR}

\begin{figure}
\centerline{\psfig{file=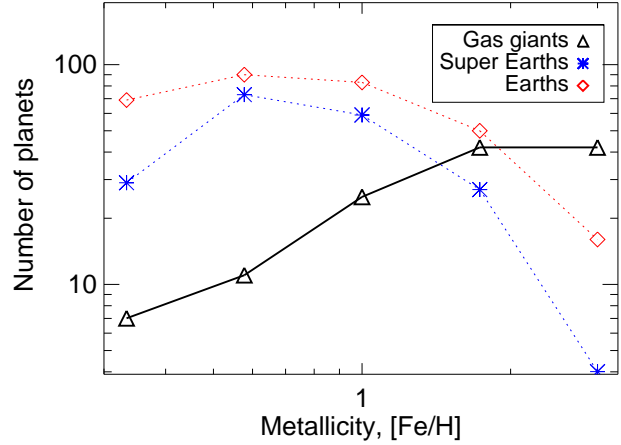,width=0.5\textwidth,angle=0}}
\caption{Number of gas giant, super-Earth and Earth-like planet groups versus
  host metallicity. Only planets left at separations $0.1 < a < 5$~AU are
  included in the figure.}
\label{fig:zcorr}
\end{figure}

Figure \ref{fig:zcorr} shows the number of planets formed in our runs at
separations $0.1$~AU~$< a < 5$~AU (this excludes the planets that reached the
innermost disc radius, e.g., the column of planets on the left-hand side of
Fig. \ref{fig:scatter}) versus host metallicity in Solar units, $z/z_\odot$,
for three groups of planets: the gas giants, the Super-Earths and the
``Earths''. The division on the three groups is made based purely on the
planet's mass. Gas giants are fragments that collapsed rather than were
disrupted, so they are all more massive than $0.5 \mj$. Super-Earths are solid
cores more massive than $5\mearth$, and Earths are planets with mass $1\mearth
< M_p < 3 \mearth$.

There is a clear and strong correlation with $z$ for gas giant planets,
although somewhat less strong than found in \cite{Nayakshin15b}. As explained
in the latter publication and in \cite{Nayakshin15a}, physically this
correlation arises from gas giants collapsing faster at higher abundance {\em
  of pebbles} in the outer disc, since the fragments then accrete pebbles at
higher rates. A larger fraction of gas clumps thus collapses before it is
tidally disrupted.

The positive correlation with $z$ in Fig. \ref{fig:zcorr} saturates at the
highest metallicity bin. It is worth emphasising that we assumed here that the
fragment birth locations do not depend on metallicity of the disc for
simplicity. Higher $z$ mean higher dust opacity, and hence longer cooling
times in the disc. One may expect the clumps to be born farther away on
average at higher $z$ due to higher disc opacity \citep[such effect is
  actually found in simulations by][]{MeruBate10b}. The farther away the
clumps are born, the longer it takes for them to migrate in, and hence the
more likely they are to survive. This secondary effect, not included in the
present simulations, may strengthen the giant planet frequency versus
metallicity trend of the model.

On the other hand, Fig. \ref{fig:zcorr} shows, surprisingly at the first
glance, no clear correlation between the host metallicity and the populations
of Earths or Super-Earths. Naively speaking, one may expect that a higher
abundance of metals {\em within the fragments} at higher metallicities would
cause more massive cores to be built. When some of the fragments are
disrupted, these cores should become visible to the observer, so this would
predict more massive cores around stars of higher metallicity.

This simple argument does not actually work because the fragment's properties
also vary with metallicity and that should be also taken into account. In
particular, a fragment loaded with pebbles at a higher rate indeed has a
higher metallicity, so that the core's accretion rate is higher, but the time
window for grains to sediment down into the core is shorter because the
fragment heats up too quickly and the grains are eventually vaporised which
clearly stops accretion of the core.

This physics is internal to the fragment and has nothing to do with the
fragment-disc interaction. Therefore, to investigate it in greater detail, we
perform a series of runs in which we turn off fragment's migration and keep
the fragment's pebble accretion rate constant, parameterised, as in
\cite{Nayakshin15a}, in terms of the "metal loading time", $t_z$, defined by
\begin{equation}
{d M_z\over dt} = {z_\odot M_p\over t_z}\;,
\label{tz_def}
\end{equation}
where $z_\odot = 0.015$ is the Solar metallicity. Thus $t_z$ is the time scale
on which the planet's metal content increases by the amount contained in that
planet at Solar metallicity. Since the fragment does not migrate it is not
tidally disrupted in these isolated runs.

\begin{figure}
\centerline{\psfig{file=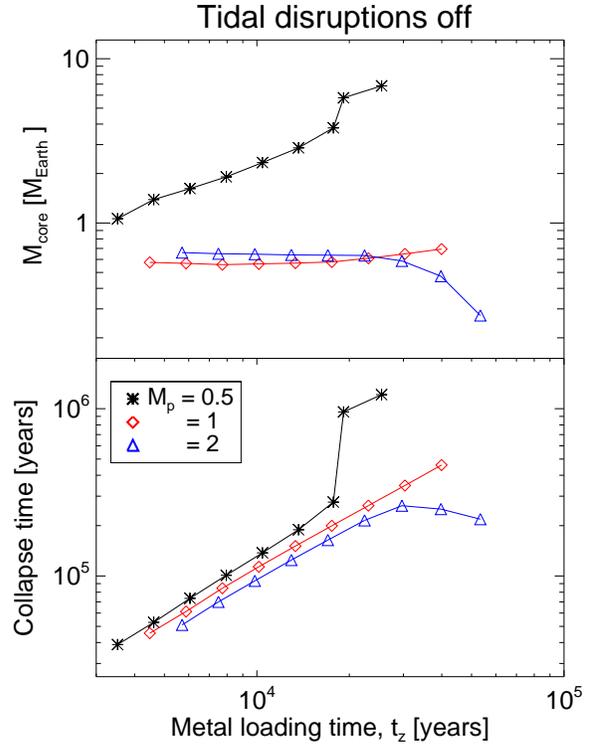,width=0.45\textwidth,angle=0}}
\caption{{\bf Top}: the mass of the solid core assembled inside the
  precollapse gas fragment as a function of the metal loading time, $t_z$, for
  non-migrating fragments of three different masses, as indicated in the
  legend. {\bf Bottom}: The time it takes the fragments to contract from the
  initial state and collapse via H$_2$ dissociation, as a function of $t_z$.}
\label{fig:Mcores}
\end{figure}

Figure \ref{fig:Mcores} shows the results of this series of runs for three
initial planet masses, $M_p = 0.5$, 1 and $2 \mj$. The top panel shows the
mass of the core assembled inside the fragment by the time it collapses versus
the metal loading time on the horizontal axis. For $M_p =1 \mj$, the core mass
is essentially independent of $t_z$. The bottom panel of Fig. \ref{fig:Mcores}
explains the reason for this finding to some degree. The panel shows the time
it took the fragment to collapse. Unsurprisingly, the faster the metals are
added to the planet, the faster it collapses, so that the core indeed has a
shorter time window to grow. For $M_p = 1\mj$ the two trends -- the faster
core growth and the shorter time available for that -- nearly cancel each
other out, yielding a nearly constant result. For $M_p=2\mj$, the dependence
is mostly flat, but at the longest $t_z$ the core mass actually
decreases. This is correlated with the time it takes for the fragment to
collapse (the bottom panel of Fig. \ref{fig:Mcores}). Physically, for $M_p =
2\mj$, the longest $t_z$ planets collapse faster because their contraction is
dominated by radiative cooling rather than pebble accretion. Radiative
contraction is faster at lower metallicities \citep{HB11,Nayakshin15a}, and
hence the result. Finally, for the lowest mass planets in
Fig. \ref{fig:Mcores}, the faster the metals are loaded into the planets (the
shorter $t_z$), the lower the mass of the cores assembled.

The different behaviour shown by the planets of different initial mass in their
response to a variance in $t_z$ is due to a fact that a number of factors,
rather than just one, control the core's assembly rate inside the planet.
The most important of these factors are: the central density of the planet,
its metal abundance, gas temperature, and the core's luminosity as this
affects the convective flux and the convective grain mixing which opposes
grain sedimentation.

The weak dependence of the core's mass on the metal loading time (and hence on
the metallicity of the disc), shown in Fig. \ref{fig:Mcores}, explains why the
populations of the rocky planets in Fig. \ref{fig:zcorr} are broadly
distributed over metallicities. Indeed, if a roughly constant fraction of
precollapse fragments behaving like those in the isolated simulations used for
\ref{fig:Mcores} is tidally disrupted, then there would be cores of similar
masses at a range of metallicities.

However, note that there is a precipitous fall in the number of rocky planets
at the highest metallicities in Fig. \ref{fig:zcorr} that would not be
explained in this picture. We believe that this decline in the number of rocky
cores with $z$ at high $z$ has to do with a changing fraction of precollapse
fragments experiencing tidal disruptions. The higher the metallicity of the
disc, the less likely the fragment is to be disrupted. This can be seen from
the bottom panel of Fig. \ref{fig:Mcores}. The fragment collapse times are
quite short ($\simlt 10^5$~years) at the low $t_z$ end. The fragment migration
times are longer than that for most of our runs in the grid of models, and
hence most of these fragments are expected to survive the tidal disruptions.

This is why at the highest metallicities {\em most} of our precollapse
fragments manage to collapse and either survive as a gas giant at the end of
the simulation or reach the inner boundary of the disc. Either way, their
rocky cores are never released from under the massive gas envelope, and this
is why there are so few rocky planets at the highest metallicities in
Fig. \ref{fig:zcorr}. This fall is probably too strong compared with
observations. We come back to the issue in the Discussion section.

\subsection{Large separations}\label{sec:largeR}

\begin{figure}
\centerline{\psfig{file=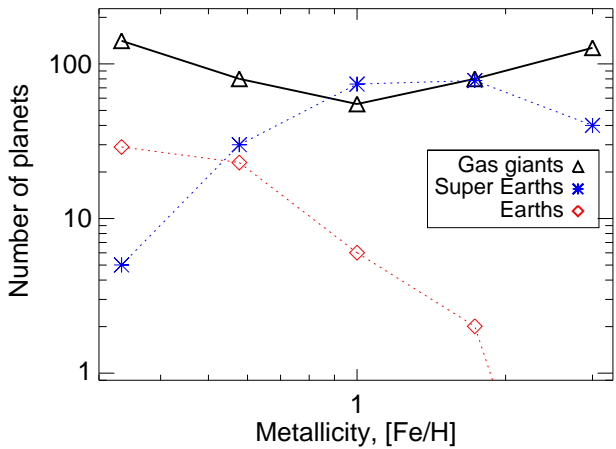,width=0.5\textwidth,angle=0}}
\caption{Number of gas giant, super-Earth and Earth-like planet groups versus
  host metallicity. Only planets found at separations $a > 10$~AU are included
  in the figure.}
\label{fig:zcorr_outerR}
\end{figure}

Figure \ref{fig:zcorr_outerR} shows the metallicity correlations for planets
survived the disc migration phase at the outer $a > 10$~AU region of the
star. For the Jovian mass planets in the figure, the planets that are stranded
far from the host star are those that migrate the slowest. Therefore, these
fragments are not likely to be threatened by the tidal disruption, so can
take as long as they need to contract. This is why the population of gas
giants is now broadly distributed over the host star metallicity, with just a
shallow minimum at around the Solar metallicity. This is quite different from
Fig. \ref{fig:zcorr} where a strong positive corelation with metallicity was
present.

The Super Earth and the Earth-like planets in Fig. \ref{fig:zcorr_outerR} are
the products of tidal disruptions of precollapse gas fragments, so these tend
to come from those fragments that were migrating faster (physically "released
to migrate" earlier, when their discs were more massive) than the gas giant
planets in the figure. The metal correlation of the population of the
Super-Earth planets in the outer 10~AU is not actually that different from
that in Fig. \ref{fig:zcorr}, with similar physics driving it.

Earth-like planets are however much less abundant at high separations from the
host, and especially at the highest metallicity bins. This is due to the fact
that very few high metallicity fragments are disrupted in the outer disc. When
they are disrupted, they tend to contain more massive cores because these
cores had more time to grow by grain sedimentation than cores disrupted in the
inner disc (which would usually migrate in much faster). This is another
interesting example where predictions of TD do not follow the simple "more
metals more cores" logic. Hopefully this prediction may be observationally
tested, although these relatively low mass planets are not likely to be found
by direct imaging surveys in near future due to their exceedingly low
luminosities.

\section{Discussion}\label{sec:discussion}

In this paper we argued for interpreting the metallicity correlations of
observed exoplanets in the framework of the Tidal Downsizing hypothesis for
planet formation rather than in the context of the Core Accretion model. In
the latter theory, the fact that massive cores form at all metallicities but
gas giants prefer metal-rich environments is odd. Once a massive core is
assembled, accreting a massive gas envelope on the top of that core is
actually easier in metal-poor environments. Lower opacities in the envelope
allow the heat of the core's assembly to escape from the gas envelope faster,
accelerating its contraction and the eventual collapse. This can be seen from
the well known result that the critical core mass for atmosphere collapse in
the CA theory increases (although weakly) with opacity of the envelope
\citep[e.g.,][]{Stevenson82,IkomaEtal00}. So, if massive cores are equally
abundant at all metallicities, one would then expect more gas giant planets at
low metallicities than at high metallicities, which is observationally not at
all the case.

In contrast, formation of gas giants in the TD hypothesis is not
preconditioned by the assembly of solid cores. As we found here, there is
actually no simple relationship between these two types of planets in
general. Both the rocky core dominated planets and the gas giants originate
from the same source -- the precollapse H$_2$ dominated fragments born by the
gravitational instability of the outer massive disc -- but the efficiency with
which the different kinds of planets are assembled from this raw material is a
function of a number of factors.

Our numerical population synthesis models allowed us to study these issues in
some detail in this paper. Focusing first on the planets surviving the disc
migration phase in the inner 5 AU of the disc (\S \ref{sec:smallR}), we found
that increasing metallicity of the environment leads to a more abundant
formation of gas giant planets, as earlier found by \cite{Nayakshin15b} for
{\em coreless} gas fragments. This result is driven entirely by the pebble
accretion accelerating collapse of the fragments at high metallicites and
helping more of them to survive.

However, formation of solid cores in our models does not correlate in a
monotonic way with the metal content of the parent discs for either Earth-like
or Super Earth planets. There are two primary reasons for this
outcome. Firstly, higher pebble accretion rates on a gas fragment do not
necessarily make more massive cores. The core's mass is equal to the integral
of the core's accretion rate over the duration of time period when the core
can grow, e.g., $\int_0^{t_{\rm grow}} \dot M_c dt$. The core's accretion
rate, $\dot M_c$, increases with increasing pebble accretion rate, but the
duration of the core assembly phase, $t_{\rm grow}$, decreases.  For these
reasons the core's mass turns out to not depend sensitively on the pebble
accretion rate (which is proxy for the disc metallicity), as could be naively
expected (see Fig. \ref{fig:Mcores}).

Secondly, in accord with the explanation offered for the observed gas giant
planet correlation with the metal abundance in TD, fewer precollapse fragments
are disrupted by the stellar tidal forces at higher metallicities. Thus, not
only the cores are not necessarily much more massive at higher metallicities,
but they are also less likely to be released from inside the overlying massive
gas envelopes.  This is the reason why the population of the rocky cores
nose-dives at the highest metallicity bin in Fig. \ref{fig:zcorr}. This
particular result is however preliminary and may weaken if we are to consider
earlier generation of gas fragments that may migrate more rapidly, so that a
sizable fraction of precollapse gas fragments is tidally disrupted even at the
highest metal abundances (which is not currently the case in our models). In
that case there will probably be more rocky planets at the high metallicity
end in Fig. \ref{fig:zcorr}.

Finally, in \S \ref{sec:largeR} we compared the metallicity correlations of
the simulated planets in the outer disc ($a > 10$~AU) to that found in the
inner disc. Our models predict that gas giant planets of all metallicites may
survive at these "cold" regions of the disc, since tidal disruptions of
precollapse fragments are far less likely there. One interesting coincidence
here is the fact that the best well known system of giant gas exoplanets
probably formed by GI of the disc, HR 8799 \citep[e.g.,][]{MaroisEtal08}, is
metal poor ([Fe/H] $\approx -0.5$).

\section{Conclusions}\label{sec:conclusions}

We believe the observed metallicity correlations of planets as a function of
their size/radius give us a valuable clue as to how planets form. It is highly
significant that gas giant planets are found almost always around metal-rich
stars \citep{Gonzalez99,FischerValenti05} but planets smaller than $\sim 4
R_\oplus$ or less massive than Neptune are abundant around stars of any
metallicity \citep{SousaEtal08,BuchhaveEtal12}.

If solid cores are assembled first as a prerequisite to gas giant planet
formation, then it is hard to see why there are massive cores in low
metallicity environments but no gas giants since the lower the opacity of the
envelope the easier it is to convert a massive core into a gas giant.

On the other hand, if gas fragments are the nurseries of massive solid cores
\citep{HS08,Nayakshin10b}, and if the fragments are preferentially destroyed
in metal-poor environments \citep{Nayakshin15a}, then the observed metallicity
correlations are to be expected as long as the fragments are able to hatch
massive cores before they are tidally disrupted. The population synthesis
models presented here appear to be consistent with the observations. They also
predict that gas giant planets and Super Earths may be abundant at large
separations from the star at all metallicities.

\section*{Acknowledgments}

Theoretical astrophysics research in Leicester is supported by an STFC
grant. This paper used the DiRAC Complexity system, operated by the University
of Leicester, which forms part of the STFC DiRAC HPC Facility
(www.dirac.ac.uk). This equipment is funded by a BIS National E-Infrastructure
capital grant ST/K000373/1 and DiRAC Operations grant ST/K0003259/1. DiRAC is
part of the UK National E-Infrastructure.

\label{lastpage}

\end{document}